\documentclass[a4paper, 12pt]{article}

\usepackage[T1]{fontenc}
\usepackage{lmodern}

\usepackage{setspace}
\usepackage{pdflscape}
\usepackage{longtable}
\usepackage{wasysym}
\usepackage{graphicx}
\usepackage{hyperref}
\hypersetup{
    colorlinks=true,
    linkcolor=blue,
    urlcolor=blue,
    citecolor=blue}
\usepackage{float}
\usepackage[
backend=biber,
style=alphabetic,
]{biblatex}
\begin{document}

\baselineskip 12pt

\begin{center}
\textbf{\Large The Predictive Brain: \\ Neural Correlates of Word Expectancy Align with \\ Large Language Model Prediction Probabilities} 

\vspace{1.5cc}
{ \sc Nikola Kölbl$^{1,2}$, Konstantin Tziridis$^{1}$, Andreas Maier$^3$, Thomas Kinfe$^4$, Ricardo Chavarriaga$^{5}$, Achim Schilling$^{1,2,4,*}$, Patrick Krauss$^{1,2,*}$}\\

\vspace{0.3 cm}

{\small $^{1}$Neuroscience Lab, University Hospital Erlangen, Germany\\ $^{2}$CCN Group, Pattern Recognition Lab, FAU Erlangen-Nürnberg, Germany\\ $^{3}$Pattern Recognition Lab, FAU Erlangen-Nürnberg, Germany\\$^{4}$Neuromodulation and Neuroprosthetics, University Hospital Mannheim, University Heidelberg, Germany\\$^{5}$ZHAW Zürich, Switzerland \\$^{*}$both authors contributed equally}
 \end{center}
\vspace{1.5cc}

\begin{abstract}
  \noindent 
 Predictive coding theory suggests that the brain continuously anticipates upcoming words to optimize language processing, but the neural mechanisms remain unclear, particularly in naturalistic speech. Here, we simultaneously recorded EEG and MEG data from 29 participants while they listened to an audio book and assigned predictability scores to nouns using the BERT language model. Our results show that higher predictability is associated with reduced neural responses during word recognition, as reflected in lower N400 amplitudes, and with increased anticipatory activity before word onset. EEG data revealed increased pre-activation in left fronto-temporal regions, while MEG showed a tendency for greater sensorimotor engagement in response to low-predictability words, suggesting a possible motor-related component to linguistic anticipation. These findings provide new evidence that the brain dynamically integrates top-down predictions with bottom-up sensory input to facilitate language comprehension. To our knowledge, this is the first study to demonstrate these effects using naturalistic speech stimuli, bridging computational language models with neurophysiological data. Our findings provide novel insights for cognitive computational neuroscience, advancing the understanding of predictive processing in language and inspiring the development of neuroscience-inspired AI. Future research should explore the role of prediction and sensory precision in shaping neural responses and further refine models of language processing.

\vspace{0.95cc}
\parbox{24cc}{{\it Language; Transformer; Large Language Models; MEG; EEG; BERT; N400; predictive coding}:
}
\end{abstract}
\onehalfspacing

\section{Introduction}
The human brain is a prediction machine, constantly anticipating upcoming sensory inputs, words, events, and in general future states \cite{friston2009predictive, smith2021recent, schilling2023predictive}. In language processing, this predictive mechanism enables fast and efficient comprehension by minimizing surprise \cite{wurm2024surprise}. When predictions fail, the brain updates its internal model to better match the incoming input, ensuring adaptive and flexible processing \cite{schilling2024bayesian, friston2009predictive}. Although predictive coding theories are well established in perception, how this framework applies to language, particularly semantic processing, remains largely unknown \cite{caucheteux2023evidence}. Despite decades of research, the neural mechanisms underlying the extraction and representation of meaning are still not fully understood \cite{pulvermuller2013neurons}.

Understanding how the brain efficiently processes language is not only a fundamental question in cognitive neuroscience, but could also provide insights for improving artificial intelligence (AI) \cite{hassabis2017neuroscience}. In recent years, transformer-based language models such as BERT, Llama and GPT-4o have revolutionized natural language processing (NLP) by using context to predict upcoming or masked words \cite{touvron2023llama, touvron2023llama2, llama32modelcard, krauss2024analyzing, kenton2019bert, Wolf_Transformers_State-of-the-Art_Natural_2020, ramezani2024analysis, kissane2024analysis, openai2022chatgpt}. These models provide a computational framework that potentially can parallel how the human brain processes linguistic input \cite{kriegeskorte2018cognitive, alkhamissi2024llm, costa2024llm, schilling2022intrinsic}. However, it remains unclear whether and to what extent neural responses in natural language comprehension reflect such statistical predictability. Investigating this relationship could bridge the gap between biological and AI, and shed light on common principles of efficient information processing \cite{hassabis2017neuroscience, chen2022far, stoll2024coincidence}.

To unravel the mechanisms of language prediction in the brain, both experimental neuroscience and computational modeling approaches are needed (see e.g. \cite{surendra2023word, stoewer2022neural, stoewer2023neural}). Advances in neuroimaging, particularly EEG and MEG, allow the tracking of brain responses to linguistic stimuli with high temporal resolution. In particular, the N400 component - a well-established neural marker of semantic processing - has been linked to predictability and surprise effects \cite{michaelov2024strong, michaelov2022more, frank2013word}. However, many studies investigation predictability with neuroimaging techniques use artificial language stimuli such as isolated words, constructions or sentences \cite{grisoni2017neural, grisoni2024predictive}, despite the fact that these experiments do not generalize well across different stimulus protocols \cite{beres2017time}. Thus, nowadays these artificial stimuli are replaced by natural stimuli and continuous speech such as audio books \cite{schilling2021analysis, koelbl2023adaptive, koelbl2024analyzing, kolbl2024methodological, garibyan2022neural, schuller2023attentional, schuller2024early}. 

In the present study, we used EEG/MEG measurements of 29 participants stimulated with a German audio book and compared event-related fields (ERF) and event-related potentials (ERP) with predictions of the BERT language model to test two main hypotheses (see \cite{koelbl2024analyzing, krauss2024temporal}). First, we hypothesized that when a word is highly predictable within continuous speech, the neural response associated with its processing - particularly in the N400 time window - should show reduced amplitude compared to less predictable words (analogously to \cite{michaelov2022more}, but using a continuous audio book rather than controlled, visually presented sentences). The second hypothesis states that, if predictive mechanisms shape language comprehension, neural activity reflecting anticipation should be more pronounced for highly predictable words even before their onset (analogously to \cite{grisoni2021correlated}).

We examined the relationship between BERT-based predictability scores and neural activity, focusing on both pre-word onset signals and N400 responses. Our results showed that as word predictability increased and consequently surprisal was reduced, N400 amplitudes decreased, in line with predictive coding theories. 
Moreover, highly predictable words indeed elicited stronger pre-activation of neural activity, indicating that the brain actively engages in anticipatory processing when predictability is high — a finding consistent with previous studies on predictive coding in language comprehension \cite{grisoni2017neural, pulvermuller2020semantic, michaelov2022more}.

\section{Methods} 

\subsection{Data}
As a natural language stimulus, we presented participants with approximately 50 minutes of the science fiction audio book Vakuum by Phillip P. Peterson, narrated by Uve Teschner (Argon Hörbuch). The audio book has several story lines, two of which were selected and divided into eight alternating chapters of approximately seven minutes each. To maintain engagement and assess comprehension, participants answered three multiple choice questions after each chapter. 

We recorded brain activity of 29 participants (15 \female, mean age: 22.8 ± 3 years) while they listened to the audio book. All participants were right-handed, native German speakers with normal hearing, no history of neurological disorders, and no use of substances. Neural responses were captured simultaneously using 248-channel magnetoencephalography (MEG) (Magnes 3600WH, 4D-Neuroimaging) and 64-channel electroencephalography (EEG) (ANT Neuro) with additional electrooculogram (EOG) and electrocardiogram (ECG) (MEG: sampling frequency = 1017.25\,Hz, EEG: sampling frequency = 2000.0\,Hz). To prevent interference from electronic components, the audio signal was delivered via an air tube from external loudspeakers into the MEG chamber, where it was played through headphones. Volume levels were adjusted individually to ensure optimal intelligibility and comfort. To minimize eye and muscle artifacts, participants were instructed to fixate on a central cross and remain as still as possible while lying down. The study was approved by the Ethics Committee of the University Hospital Erlangen (Approval No: 22-361-2, PK).

\subsection{Predictability Scores}\label{PredictabilityScores}

\begin{figure}[!htb]
\centering
    \includegraphics[width=\textwidth]{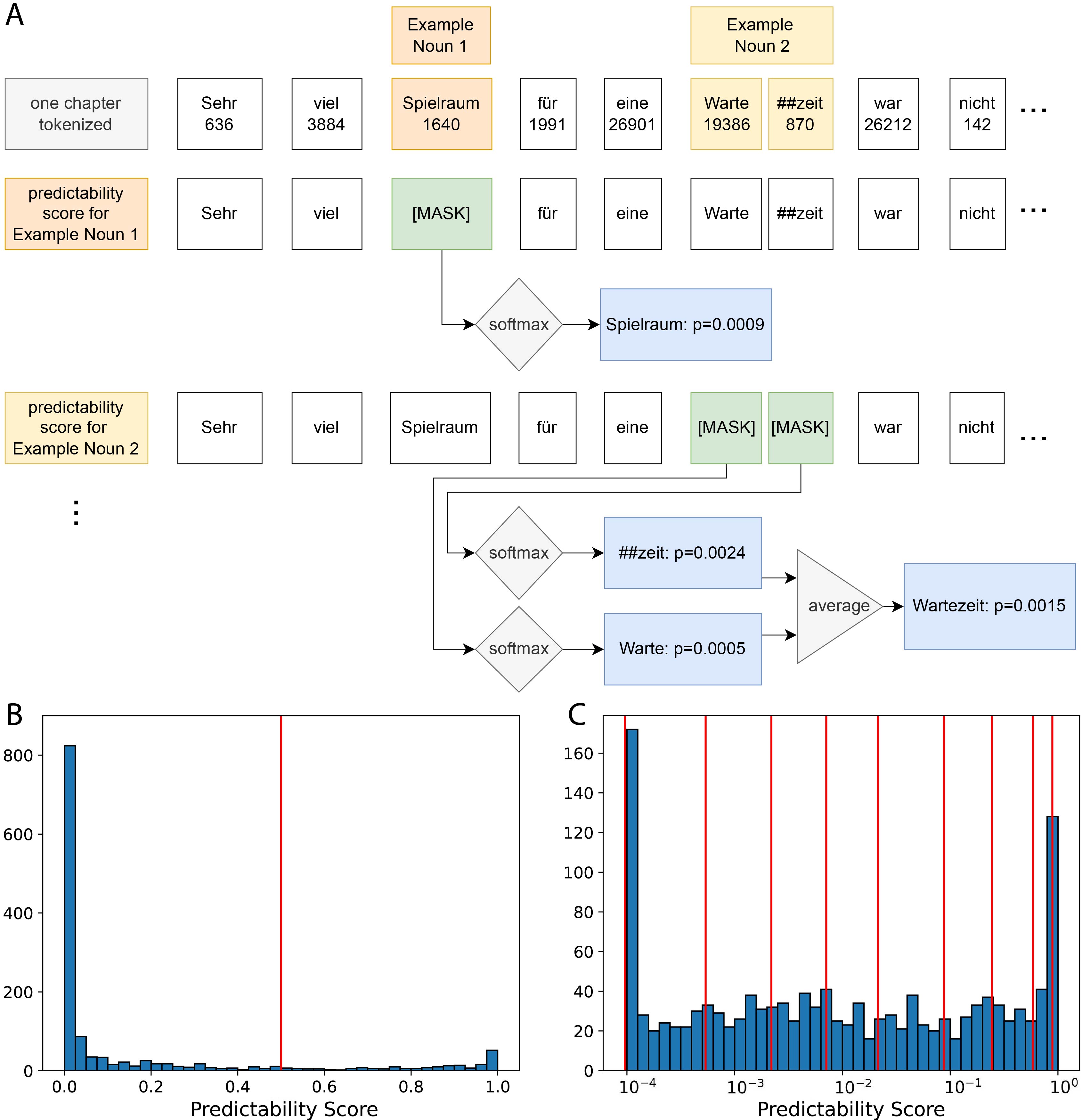}
    \caption{(A): Scheme for computing predictability scores for individual nouns in the audio book. Each target noun (and its corresponding sub-words) is masked ([MASK]), and the resulting BERT output is processed through a softmax function to determine the probability of the original word. For multi-sub-word tokens, the final predictability score is obtained by averaging the probabilities across all sub-words. \\(B): Histogram of predictability scores of all nouns (red vertical line: predictability of 0.5 used to divide low and high predictable words). (C): Semi-logarithmic plot of the histogram including thresholds for dividing scores into 10 equal splits used for correlation analysis (red lines).}
    \label{histo}
\end{figure}

To quantify the predictability of each noun in the text - the likelihood of its occurrence in a given context - we used BERT (Bidirectional Encoder Representations from Transformers), a transformer-based language model. BERT processes words bidirectionally, capturing contextual dependencies across entire sentences \cite{kenton2019bert}. For our analysis, we used the pre-trained German model 'bert-base-german-cased' via the Hugging Face Transformers library \cite{Wolf_Transformers_State-of-the-Art_Natural_2020}.
The text was first pre-processed by removing all punctuation. Next, each word token (along with any associated sub-tokens) was iteratively masked - one at a time - before being passed to BERT for prediction. To ensure that the model remained in inference mode, gradient computations were disabled using PyTorch, as we were not interested in fine-tuning the model \cite{NEURIPS2019_9015}. The modified text containing the masked token was then fed into BERT, and the model's output logits were normalized using a softmax function to obtain probability values between 0 and 1 for each word (see Fig. \ref{histo} A for process diagram).
For words that were split into multiple sub-tokens, the final predictability score was calculated as the average probability across all sub-word-components (Fig. \ref{histo} B). This process was applied to all the words in the audio book, after which we extracted predictability scores only for nouns.

The distribution of predictability scores for all nouns in the audio book shows a clear imbalance, with the majority of nouns having lower predictability. To facilitate the analysis, the predictability scores were binned into two categories: low predictability (scores < 0.5) and high predictability (scores > 0.5, see Fig. \ref{histo} B, threshold in red). A significant class imbalance was observed, with 1,182 trials in the low predictability category compared to only 194 trials in the high predictability group. To ensure a balanced comparison, a subset of low-predictability trials was randomly selected to match the number of high-predictability trials, thus creating an equal distribution across conditions. For further analysis, we applied a semi-logarithmic scaling approach, correlating neural responses with the logarithm of predictability scores (Fig. \ref{histo} C). This transformation reverses the softmax operation of predictability scores and allows for a more refined assessment of the relationship between linguistic predictability (probabilities) and brain activity. To explore potential systematic patterns in brain activity correlated with predictability scores while maintaining consistent signal-to-noise ratios across categories, we divided the scores into 10 equal-sized bins, each containing approximately 137 trials (Fig. \ref{histo} C). 

\subsection{Data Preparation}
To improve the quality of the EEG and MEG data, we applied a standard pre-processing pipeline \cite{ferrante2022flux}. Data processing was performed using MNE-Python (v1.8.0), starting with the identification and interpolation of bad sensors and electrodes, defined as those with flat or excessively noisy signals using Maxwell filtering \cite{gramfort2013meg}. We then applied a 1-20\,Hz bandpass filter to remove irrelevant frequency components. For computational efficiency, the data was downsampled to 200\,Hz before performing independent component analysis (ICA) for artifact rejection. To remove artifacts related to eye movements and cardiac activity, we excluded the first two independent components (ICs) with the highest variance, along with any additional ICs that correlated with simultaneously recorded electrooculogram (EOG) or electrocardiogram (ECG) signals \cite{koelbl2023adaptive}.

To segment the continuous MEG signal, we employed Forced Alignment using \textit{WebMAUS} \cite{schiel2015statistical}, aligning the audio files with their corresponding transcripts to extract precise word onset timestamps from the audio book. The audio signal was simultaneously recorded on a separate stimulus channel during playback, allowing for precise synchronization. Using these word onset markers, we segmented MEG and EEG data into epochs spanning from -1.0 to +2.0\,s relative to word onset, with baseline correction applied from -1.0 to 0.0\,s.
Since our primary focus was on the differential processing of highly-predictable vs. low-predictable nouns, we used the Natural Language Processing (NLP) software spaCy (model: de\_core\_news\_sm) to classify words in the audio book by their part-of-speech tags \cite{Honnibal_spaCy_Industrial-strength_Natural_2020}. We analyzed ERPs and ERFs for nouns and identified peak responses in the time window from -1.0\,s until 2.0\,s and extracted the topographic distribution at the time point of the strongest negative responses (Figure \ref{amplitudes} A and B). For EEG data, we selected parietal electrodes: CP2, CPz, CP1, P2, Pz, P1. For MEG data, we focused on left frontal sensors: A229, A212, A178, A154, A126, A230, A213, A179, A155, A127, A177, A153, A125. We divided the nouns first into low and high predictable and then into the ten bins mentioned in Chapter \ref{PredictabilityScores} and calculated corresponding ERPs and ERFs.

\subsection{Statistical Tests}
To identify time windows in which low- and high-predictable nouns elicited significantly different neural responses, we performed a paired Wilcoxon signed-rank test with 5,000 randomizations with Brainstorm \cite{pantazis2005comparison, tadel2011brainstorm}. A false discovery rate (FDR) correction was applied across signal, time and frequency dimensions to control for multiple comparisons.  \\
For calculating correlations we extracted the mean activities in the significant time intervals over the selected channels and sensors for each of the ten binned nouns ERPs/ERFs. We then fit a linear regression on these values using sklearn.linear\_model and calculated the p-values and r$^2$-values using the libraries numpy and scipy to quantify the relationship between predictability and brain activity \cite{scikit-learn, harris2020array, 2020SciPy-NMeth}.

\subsection{Source Reconstruction}
For source reconstruction, we used the open source software Brainstorm \cite{tadel2011brainstorm} with the standard ICBM 152 brain anatomy from the MNI database to generate a head model \cite{fonov2009unbiased}. For MEG data we used the overlapping spheres method to compute a cortical surface head model, while for EEG data we used a boundary element model (BEM) with OpenMEEG \cite{huang1999sensor, gramfort2010openmeeg}. The noise covariance was estimated from a one minute silent baseline recorded immediately before the audio book was played. For source estimation, we used minimum norm imaging with the sLORETA method and constraint dipole orientations \cite{pascual2002standardized}. Source reconstructions were first computed for participant-level averages of low- and high-predictability noun ERPs and ERFs, followed by a grand average across subjects.

\section{Results} 
\subsection{ERF/ERP Analysis}

\begin{figure}[!htb]
    \centering
    \includegraphics[width=\textwidth]{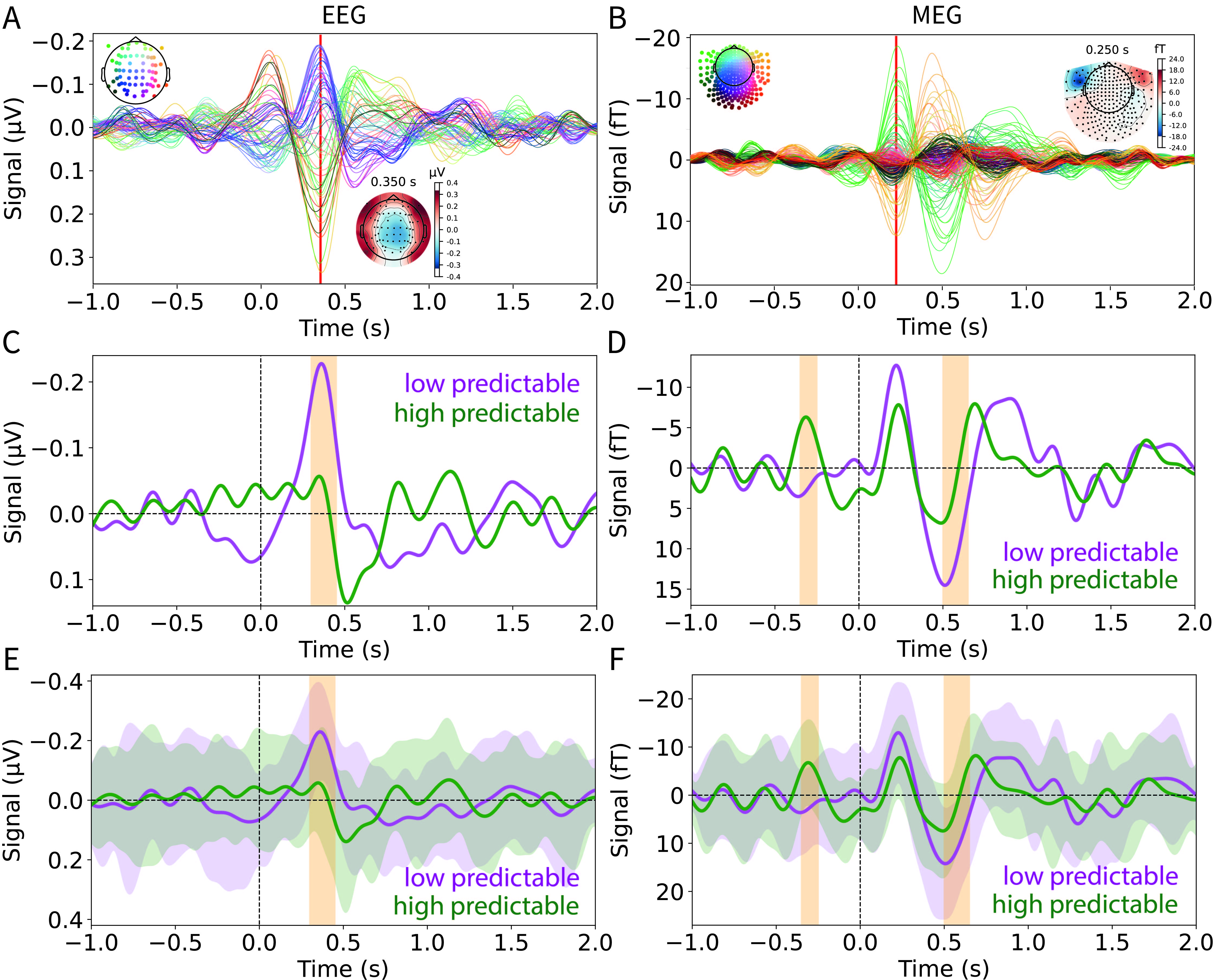}
    \caption{A: Grand average nouns ERP (EEG) including topographic map at peak 350\,ms after word onsets. (B): Grand average nouns ERF (MEG) with topographic map at 250\,ms after word onsets.
    C, D: Comparison of ERPs and ERFs of high (green) and low (purple) predictable nouns; ERPs averaged across parietal channels (CP2, CPz, CP1, P2, Pz, P1) and ERFs averaged across left frontal channels (A229, A212, A178, A154, A126, A230, A213, A179, A155, A127, A177, A153, A125). Orange background colors highlight significant p-values (p$<$0.05; FDR corrected, permutation paired test statistic: Wilcoxon signed-rank test). E, F: same as C, D but with variance.}
    \label{amplitudes}
\end{figure}

To investigate the effect of predictability on neural processing, we compared brain responses to low (predictability score < 0.5) and high (predictability score > 0.5) predictable nouns (Fig. \ref{amplitudes} A, B: ERPs and ERFs for all nouns (high and low predictability); Fig. \ref{amplitudes} C, D: ERF/ERP comparison between high/low predictable nouns). Cluster-based permutation statistics were used to identify significant time windows while controlling for multiple comparisons (Fig. \ref{amplitudes} C, D) \cite{pantazis2005comparison}. In the EEG data, differences emerged 300-450\,ms after word onset, consistent with the well-established N400 component typically associated with semantic processing (Fig. \ref{amplitudes} C). In the MEG data, significant effects were observed both before word onset (-350 to -250\,ms) and after word onset (500-650\,ms). \\ To confirm that these effects were not due to chance, we analyzed two additional low predictable data subsets for both EEG and MEG. The EEG second subset showed a trend toward significance from 320 to 520\,ms (p = 0.11), while the third subset exhibited a significant effect from 310 to 460\,ms (p < 0.05). Both MEG subsets revealed consistent significant differences: subset 2 from -330 to -270\,ms, 520 to 700\,ms, and 820 to 920\,ms; subset 3 from -310 to -250\,ms, 590 to 740\,ms, and 820 to 960\,ms. These matching patterns across subsets suggest the presence of consistent and reproducible effects.\\ While the early effect suggests an anticipatory process in response to less predictable words, the later effect may reflect N400-related activity, similar to the EEG findings (Fig. \ref{amplitudes} C). In both data sets, high-predictability nouns elicited lower amplitude responses than low-predictability nouns, supporting the idea that when word expectancy is high, the brain expends less effort during processing \cite{michaelov2022more}. These results are consistent with the hypothesis that the brain is more engaged when encountering less predictable words, likely reflecting increased processing demands during lexical access and integration.

\subsection{Source Space Analysis}

\begin{figure}[!hbt]
\centering
    \includegraphics[width=\textwidth]{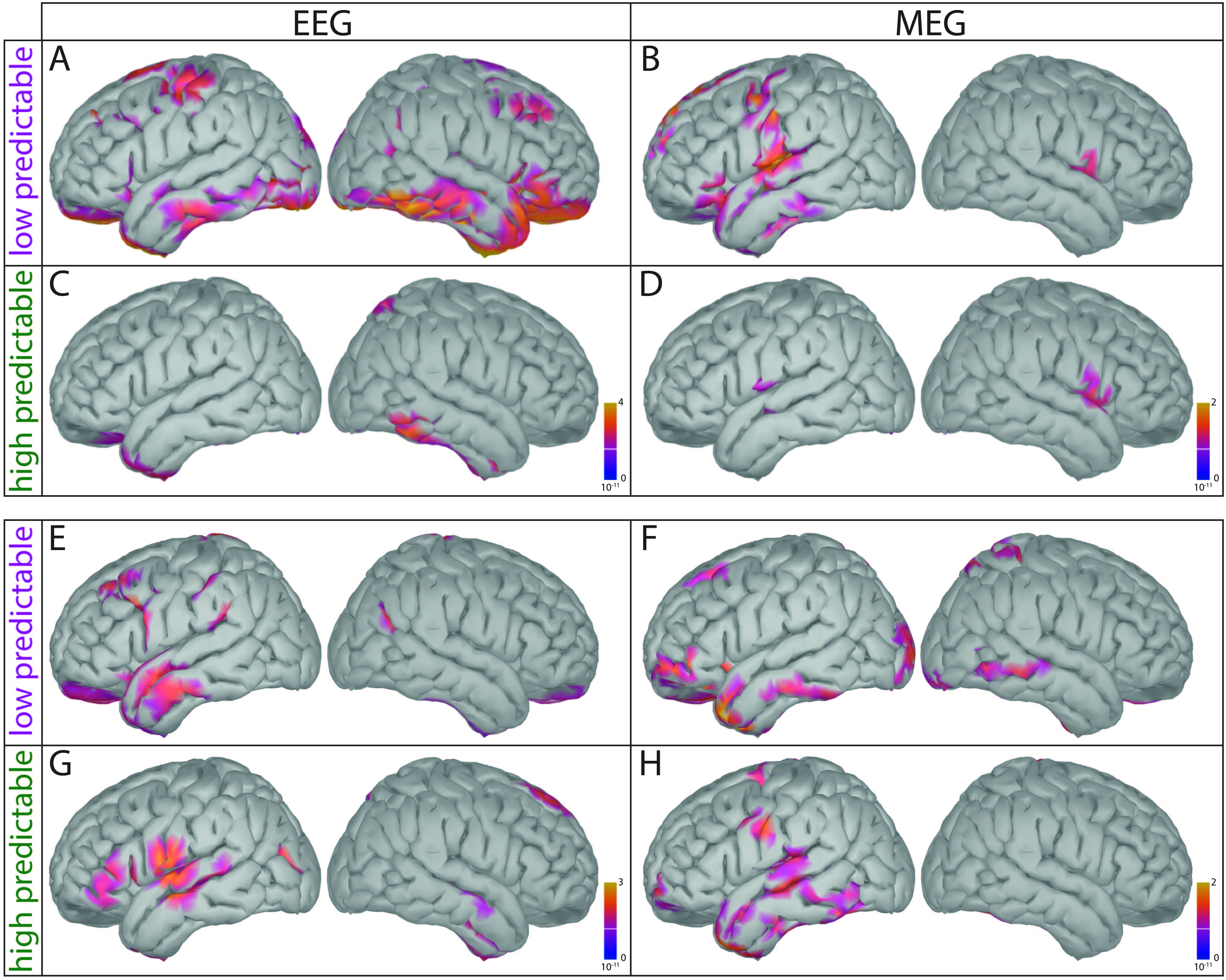}
    \caption{RMS amplitudes in source space for EEG data in the time interval 300\,ms-450\,ms for low predictable (A) and high predictable nouns (C).
    RMS amplitudes in source space for MEG data in the time interval 500\,ms-650\,ms for low predictable (B) and high predictable nouns (D); 
     E-H: Predictive activity before word onset (time intervals: -100\,ms-0\,ms for EEG and -350\,ms until -250\,ms for MEG) for EEG (E,G) and MEG (F, H) }
    \label{source}
\end{figure}

To specify the brain regions involved in semantic processing, we examined the source reconstruction of event-related fields (ERFs, Fig. \ref{source} B, D) and event-related potentials (ERPs, Fig. \ref{source} A, C) for low- and high-predictability nouns. Consistent with the observed differences in sensor-level data, source estimates revealed that low-predictability nouns elicited stronger neural responses and engaged more extensive cortical regions compared to high-predictability nouns after word onset. These effects were evident within the significant time windows identified previously, with increased activation in EEG (300-450\,ms after onset) and MEG (500-650\,ms after onset). In particular, significant activity was observed in parietal cortex and sensorimotor regions, suggesting that low-predictability nouns elicit broader cortical engagement, possibly reflecting increased processing demands for semantic integration and motor-related aspects of speech perception \cite{bonnet2022kinesthetic, tian2023spatiotemporal, pulvermuller2005functional}. The widespread recruitment of these areas supports the idea that predictability modulates the neural effort required for language comprehension, with greater activation occurring when word expectancy is low \cite{michaelov2022more}. To further investigate the anticipatory mechanisms involved in predictive language processing, we analyzed source activity prior to word onset (Fig. \ref{source} E-H). Significant differences in activation patterns between low and high predictable nouns were observed within the pre-onset time windows (-100 to 0\,ms for EEG and -350 to -250\,ms for MEG). In the EEG data, highly predictable nouns were associated with greater activation in left fronto-temporal regions (Fig. \ref{source} G), showing the expected left-hemispheric lateralization characteristic of language processing. In contrast, slightly different effects were observed in the MEG data (Fig. \ref{source} F, H); however, there was a tendency for increased pre-onset activity in sensorimotor regions for low-predictability nouns (Fig. \ref{source} F), suggesting a possible involvement of motor-related processes in linguistic anticipation (see e.g. \cite{singer2024speech}).

\subsection{Correlation Analysis}

\begin{figure}[!hbt]
    \centering
    \includegraphics[width=\textwidth]{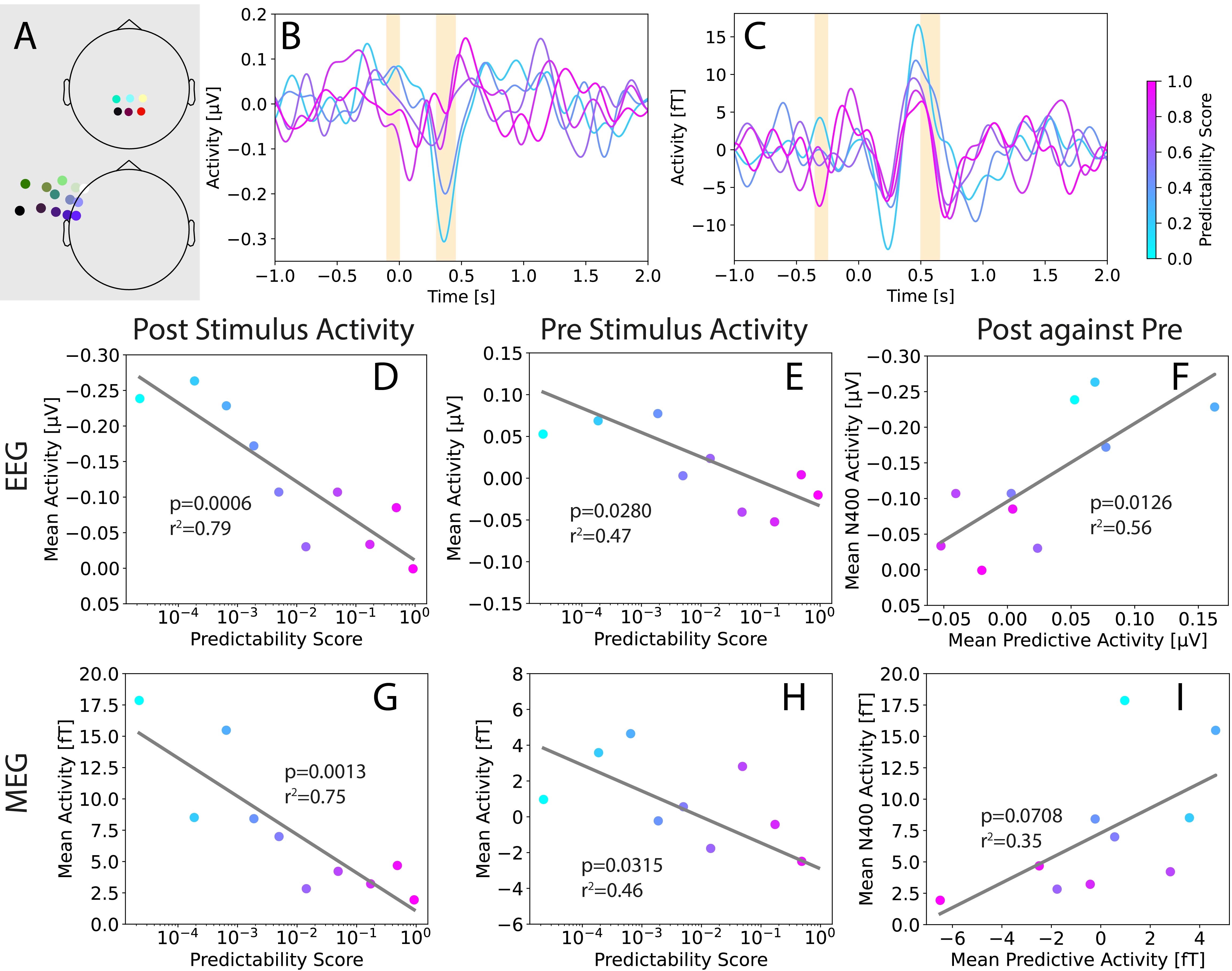}
    \caption{Correlation analysis between predictability scores of BERT and neural activity in ten independent splits. Mean activity over parietal channels (CP2, CPz, CP1, P2, Pz, P1) for EEG (A top) and over left frontal sensors (A229, A212, A178, A154, A126, A230, A213, A179, A155, A127, A177, A153, A125) for MEG (A bottom) of every second split in (B: EEG) and (C: MEG) color coded from cyan (low predictable) to magenta (high predictable). Regression analysis for post stimulus activities show high correlation with predictability scores with p=0.00006, r$^2$=0.79 for EEG data in time frame 300\,ms-450\,ms (D) and p=0.0013, r$^2$=0.75 for MEG data in time frame 500\,ms-650\,ms (G). Pre stimulus activities also reveal significant correlation for EEG (E, p=0.0280, r$^2$=0.47 in time frame -100\,ms-0\,ms) and MEG data (H, p=0.0315, r$^2$=0.46 in time frame -350\,ms to -250\,ms). Relation between mean post- and mean pre-stimulus activity show negative correlation (F for EEG data: p=0.0126, I for MEG data: p=0.0708, r$^2$=0.35).}
    \label{corr}
\end{figure}

To gain a more detailed understanding of whether the effects described above are binary or continuous, we performed a correlation analysis between predictability scores and neural activity in the "N400" time windows 300\,ms-450\,ms for EEG and 500\,ms-650\,ms for MEG after word onset. To ensure the robustness of this relationship, we examined the mean amplitudes within the identified time windows across ten independent splits of the data over parietal channels in EEG and left frontal sensors in MEG (see Fig. \ref{corr} A for channel/sensor selection, B and C for exemplary EEG and MEG signals). This approach allowed us to assess how gradual changes in predictability affected neural signal strength. By averaging the predictability scores within each split, a clear trend emerged: higher predictability was consistently associated with lower neural response amplitudes (Fig. \ref{corr} D, G). Regression analysis revealed highly significant correlations in both EEG (p = 0.0006, r$^2$ = 0.79) and MEG (p = 0.0013, r$^2$ = 0.75) data, demonstrating a systematic and robust relationship between predictability and neural response amplitude (see Fig. \ref{corr} D, G). The fact that these correlations are highly significant further supports the idea that brain activity scales continuously with linguistic predictability, rather than operating in a binary fashion. This systematic reduction in signal amplitude reinforces the hypothesis that the brain expends more effort processing less predictable words.
Building on these findings, we next investigated whether the reduction in neural activity during word processing could be driven by predictive mechanisms that occur prior to word onset. To test this, we analyzed pre-onset neural activity in both EEG and MEG (Fig. \ref{corr} E, H). For the MEG data, we focused on the previously identified significant time window (-350 to -250\,ms before word onset), while in the EEG data, where no significant pre-onset effects were observed, we examined a pre-word interval at -100\,ms to assess early anticipatory processes. Analysis revealed a significant relationship between pre-onset neural activity and predictability, although the effect was less pronounced than during word processing. Both EEG (p = 0.028, r$^2$ = 0.47) and MEG (p = 0.0315, r$^2$ = 0.46) data showed that higher predictability was associated with larger negativities. These results suggest that predictability modulates neural activity even before word onset, supporting the idea that the brain engages in anticipatory processing when the linguistic context allows for reliable predictions \cite{grisoni2024predictive}. This pre-stimulus activity is negatively correlated with the N400 amplitude (Fig. \ref{corr} F: (EEG) p = 0.0126, r$^2$ = 0.56, I: (MEG) p = 0.0708, r$^2$ = 0.35). The negative correlation between N400 amplitude and pre-stimulus activation supports predictive coding, where stronger anticipatory activity reduces processing demands at word onset, leading to a smaller N400 response when predictions match input (see also \cite{grisoni2021correlated}).

\section{Discussion}
This study investigated how word predictability modulates neural activity during natural language processing, focusing on its effects during word recognition and anticipatory processing. EEG and MEG recordings were simultaneously collected from 29 participants listening to an audio book, with predictability scores assigned to nouns using BERT. Neural responses differed significantly between nouns with low and high predictability, with EEG effects in parietal electrodes and MEG effects in left frontal sensors. Detailed analysis confirmed a significant correlation between predictability and neural activity approximately 400\,ms after word onset (N400 wave), showing that higher predictability was associated with reduced signal amplitude during word processing. We also found that pre-onset neural activity was associated with predictability, with greater fronto-temporal activation in EEG for highly predictable words. In MEG, low-predictability words showed a tendency for increased sensorimotor activity, suggesting a possible motor-related component to linguistic anticipation.  The fact that motor planning and language anticipation / production are closely linked is highlighted by the fact that it is possible to build brain-computer interfaces (BCI) to control a mouse cursor based on an implant in the ventral precentral gyrus, which is related to tongue movement and thus speech production \cite{singer2024speech}. A detailed understanding on predictive coding, language understanding, and production can also have some impact on the development of universally applicable BCIs.

Our main finding that the N400 amplitude is inversely correlated with predictability scores, derived from a language model, is consistent with previous research on semantic processing. Goldstein et al. reported increased neural activity for unpredictable words around 400\,ms after onset, consistent with the N400 effect \cite{goldstein2022shared}. Similarly, Maess et al. showed that more predictable nouns elicit a smaller N400 response compared to less predictable ones \cite{maess2016prediction}. Importantly, our study extends these findings by using a combined MEG/EEG setup with more naturalistic linguistic stimuli, demonstrating that the relationship between N400 amplitude and predictability holds even in rich, continuous language contexts.
The observed correlation between N400 amplitude and surprisal (negative log conditional probability from transformer networks \cite{michaelov2024strong}) highlights the role of predictability in semantic processing. Recent work has shown that surprisal estimates derived from large language models (LLMs) such as GPT-3 reliably predict N400 effects, strengthening the link between computational models of language prediction and neural responses \cite{michaelov2024strong, mischler2024contextual}. Similar findings have been reported by Kuperberg and coworkers \cite{kuperberg2020tale}, also based on reading and thus visual paradigms, highlighting the need to investigate whether these effects generalize to naturalistic auditory language, resp. speech processing. Beyond predictability, N400 amplitudes may also be shaped by the precision of predictions and the strength of bottom-up sensory input, reflecting the brain's confidence in a given prediction \cite{lecaignard2022neurocomputational}. This additional variability in prediction confidence provides an avenue for future research, offering a more nuanced perspective on how linguistic predictions modulate neural processing.

Our second main finding, that higher predictability is associated with greater pre-activation of neural representations prior to word onset, is consistent with previous research on anticipatory language processing. Predictable words have been shown to elicit enhanced pre-onset neural activity, likely reflecting context-driven lexical anticipation \cite{goldstein2022shared, grisoni2017neural, grisoni2024predictive}.
In this context, the so-called semantic prediction potential (SPP) has been introduced as new neural marker of predictive processing \cite{grisoni2021correlated}. We found that the SPP correlates positively with word predictability and negatively with the N400 wave, further linking pre-activation mechanisms to lexical integration. Thus, we could confirm the finding of Grisoni and co-workers, who showed similar correlations using more surrogate stimuli \cite{grisoni2021correlated}. However, further research is needed to fully understand the relationship between contextual predictability, SPP and N400. These effects are influenced by factors such as the precision (inverse variance, \cite{lecaignard2022neurocomputational}) of the prediction (prior) and bottom-up sensory input (sensory precision), which also affect the explanatory power of LLM-derived surprise scores as predictors of ERP components \cite{lecaignard2022neurocomputational, krieger2024limits}.
In particular, the role of prediction precision, which can be interpreted as a confidence score, is supported by evidence that the amplitude of N400 correlates with subjective belief states \cite{lecaignard2022neurocomputational}. Specifically, studies have shown that N400 responses to semantic violations in human-written texts are stronger than those generated by LLMs \cite{rao2024comprehending}. Furthermore, when reading texts generated by LLM, the amplitude of N400 increases when individuals have greater confidence in the competency of the model \cite{rao2024comprehending}, suggesting that belief in the reliability of the language model may influence the accuracy of the prediction \cite{lecaignard2022neurocomputational, schilling2023predictive}.

In summary, our study highlights the role of predictive coding in naturalistic language processing, emphasizing the integration of anticipatory and sensory signals. To our knowledge, this is the first study to demonstrate these effects using naturalistic speech stimuli, thereby bridging computational language models with neural measures in real-world contexts. This work advances cognitive computational neuroscience and neuroscience-inspired AI by opening new avenues for exploring the interplay between top-down predictions and bottom-up inputs \cite{kriegeskorte2018cognitive, hassabis2017neuroscience}.

\section{Acknowledgments}
This work was funded by the Deutsche Forschungsgemeinschaft (DFG, German Research Foundation): grant TZ\,100/2-1 (project number 510395418) to KT, grants KR\,5148/2-1 (project number 436456810), KR\,5148/3-1 (project number 510395418), KR\,5148/5-1 (project number 542747151), and GRK\,2839 (project number 468527017) to PK, and grant SCHI\,1482/3-1 (project number 451810794) to AS. Furthermore, the research leading to these results has received funding from the European Research Council (ERC) under the European Union’s Horizon 2020 research and innovation programmme (ERC Grant No. 810316) to AM.

\section{Author Contributions}
AS and PK developed the study protocol. AS, PK, KT, AM supervised the study. NK performed the measurements. NK developed the evaluation programs. All authors discussed the results. AS, PK, NK drafted the first version of the manuscript. AM and TK reviewed the first draft of the manuscript. All authors accept the final version of the manuscript.

\printbibliography

\end{document}